\documentclass[osajnl,twocolumn,showpacs,superscriptaddress,10pt]{revtex4-1} %% use 11pt for Applied Optics
\usepackage{amsmath,amssymb,graphicx}
\begin{document}

\title{Rotational cavity optomechanics}

\author{M. Bhattacharya}
\affiliation{School of Physics and Astronomy, Rochester Institute of Technology,
84 Lomb Memorial Drive, Rochester, NY 14623, USA}

\begin{abstract} We theoretically examine the optomechanical interaction between a rotating nanoparticle
and an orbital angular momentum-carrying optical cavity mode. Specifically, we consider a dielectric nanosphere rotating
uniformly in a ring-shaped optical potential inside a Fabry-Perot resonator. The motion of the particle is probed
by a weak angular lattice, created by introducing two additional degenerate Laguerre-Gaussian cavity modes carrying
equal and opposite orbital angular momenta. We demonstrate that the rotation frequency of the nanoparticle
is imprinted on the probe optical mode, via the Doppler shift, and thus may be sensed experimentally using homodyne
detection. We show analytically that the effect of the optical probe on the particle rotation vanishes in the regime
of linear response, resulting in an accurate frequency measurement. We also numerically characterize the degradation
of the measurement accuracy when the system is driven in the nonlinear regime. Our results are relevant to
rotational Doppler velocimetry and to studies of rotational Brownian motion in a periodic lattice.
\end{abstract}

% REPLACE WITH CORRECT OCIS CODES FOR YOUR ARTICLE
                          % NOTE: \ocis{} IS ALIASED TO \pacs{} BUT MUST
                          % FORMAT THE TERMS CORRECTLY FOR EACH JOURNAL

\maketitle %% required
\section{Introduction}
Cavity optomechanics, which realizes the coupling of optical radiation to mechanical motion, is
a versatile platform for sensing technologies \cite{Kippenberg2008,Girvin2009,Aspelmeyer2013,Meystre2013}.
Prime instances of this capability are
the ultrasensitive linear displacement detectors used for gravitational wave interferometry \cite{McClelland2011}
and atomic force microscopy \cite{Liu2012}, as well as recently developed accelerometers \cite{Krause2012},
magnetometers \cite{Forstner2012} and thermometers \cite{Barker2014}. Given its effectiveness in supporting
detection functionalities, it is relevant
to inquire if cavity optomechanics can be used for sensing nanomechanical rotation, a capability critical to fields
such as nanotechnology \cite{WangBook}, Doppler velocimetry
\cite{Parkin2004,Yoshi2013,Lavery2013,Phillips2014,Lavery2014,Guzman2014},
atomtronics \cite{Wright2013} and statistical mechanics \cite{Grier2003,Blum2006}.

In this article, we consider the rotational motion of a nanoparticle inside a Fabry-Perot cavity.
While most of cavity optomechanics has addressed vibrational motion
\cite{Aspelmeyer2013}, there is some precedence for studies of rotational motion \cite{Law2012}. Torsional
oscillators have been used to address angular deflections \cite{Tittonen1999,MBPMLin2007,Mueller2009,Isart2010,Kim2013}; but
these systems exhibit restricted, rather than free, mechanical rotation. Pseudospins such as cold atoms \cite{Brahms2010}
and Bose-Einstein condensates \cite{Jing2011} have been coupled to optical cavity fields. However, these systems do not actually
involve physical rotation; instead, their internal degrees of freedom mathematically map on to rotor models. Optomechanical gyroscopes
have been considered, but they require rotation of the optical cavity itself \cite{Norgia2001,Bhave2014,Davuluri2013}.

In a pioneering theoretical suggestion, the orbital motion of a nanoparticle around a spherical whispering gallery mode resonator was
shown to generate optical signatures of rotational motion \cite{Deych2011}. However, in that case the particle is kept in
rotation by a torque arising from the optical mode itself. This torque thus ultimately determines
the particle rotation frequency, and its presence is undesirable for sensing an arbitrary, externally imposed nanomechanical
rotation rate, or for investigating dynamics that are driven only by a thermal torque, i.e. Brownian motion. Further, the
optical torque in the earlier proposal is nonconservative, leading to complicated non-Hamiltonian dynamics. Lastly, the
nanoparticle in that case can only interact with three values of the optical angular momentum, namely $0$ and
$\pm \hbar$ \cite{Deych2011}. This constraint places a restriction on the general exploitation of the orbital angular momentum
(OAM) degree of optical freedom.

In this article, we investigate a cavity optomechanical scheme which is sensitive to full mechanical rotation
and does not require motion of the cavity. Unlike previous work, our proposal
utilizes two optical modes, a \textit{trap} mode which exerts no torque on the nanoparticle, and a \textit{probe}
mode which exerts a small (conservative) torque with negligible effect on the particle rotation in the regime of linear
response of the system. Lastly, the particle in our scheme can couple to any value of the optical OAM $l$; this
opens up the possibility of a cavity-based version of standard rotational Doppler velocimetry
\cite{Yoshi2013,Lavery2013,Phillips2014,Lavery2014}. In order to demonstrate our scheme, we characterize
theoretically below the measurement of the constant rotation rate of a uniformly orbiting nanoparticle.

\begin{figure}[b!]
\includegraphics[width=0.6\columnwidth]{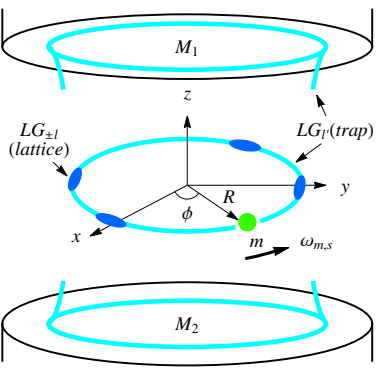}
\caption{(Color online). The optomechanical nanorotation sensing scheme proposed in this article. Cavity mirrors
$M_{1}$ and $M_{2}$ confine an intense Laguerre-Gaussian beam $LG_{l'}.$ This annular beam traps a dielectric
nanoparticle of mass $m$ on a ring of radius $R$. The dielectric rotates at a constant mechanical frequency
$\omega_{m,s}$. Its motion is probed by a weak angular lattice formed by two additional Laguerre-Gaussian
beams $LG_{\pm l}$. In the figure, $l=2$, resulting in four lattice sites on the trapping ring. The rotation of the
particle is detected by homodyning the probe mode $LG_{\pm l}$ after it exits the cavity.}
\label{fig:P1}
\end{figure}

\section{Configuration}
In this section we describe the physical setup underlying our system of interest. We first briefly address the trapping
of the nanoparticle in a cavity, and then offer a more detailed description of the detection of its rotation.
\subsection{Trap mode}
We consider a Fabry-Perot cavity with two highly reflective mirrors, as shown in Fig.~\ref{fig:P1}.
Two optical fields excite the cavity. The first, trapping, field is an intense Laguerre-Gaussian beam $LG_{l'}$
carrying orbital angular momentum $l' \neq 0$, at a wavelength $\lambda_{t}$ resonant with the cavity. The
practicality of exciting such cavity modes has been considered earlier, both theoretically \cite{Bond2011} as well as
experimentally \cite{Gatto2014}. For such a mode, the corresponding intracavity intensity is \cite{Wright2000}
\begin{equation}
I_{l'}(\vec{r})=\frac{P_{in}'}{|l'|!\mathcal{T}\pi R^{2}}\left(\frac{r\sqrt{2}}{R}\right)^{2\left|l'\right|}
e^{-2r^{2}/R^{2}}\cos^{2}\left(k_{t} z\right),
\end{equation}
where $r$ and $z$ are radial and axial cylindrical coordinates, respectively, $\mathcal{T}$ is the intensity transmission
coefficient of either cavity mirror,
\begin{equation}
k_{t}=\frac{2\pi}{\lambda_{t}},
\end{equation}
is the wave-vector, $P_{in}'$ is the input power, and
\begin{equation}
R=\omega_{0}\left(\frac{|l'|}{2}\right)^{1/2},
\end{equation}
is the mode spot size \cite{Wright2000}, $\omega_{0}(\gg \lambda_{t})$ being the beam
waist.

The energy of interaction between the trapping optical mode and a dielectric sphere of mass $m$ and radius
$r_{d} <\lambda_{t}$ (Rayleigh regime) in the dipole approximation is
\begin{equation}
H_{\mathrm{int}}(\vec{r})=-\frac{2\alpha}{c\epsilon_{0}}I_{l'}(\vec{r}),
\end{equation}
where
\begin{equation}
\alpha =3 V \epsilon_{0}\left(\frac{\epsilon_{r}-1}{\epsilon_{r}+2}\right),
\end{equation}
is the dielectric polarizability, $\epsilon_{0}$ the
permittivity of vacuum, $\epsilon_{r}$ the relative dielectric constant, $c$ the velocity of light and
\begin{equation}
V=\frac{4\pi r_{d}^{3}}{3},
\end{equation}
the volume of the dielectric sphere. Here we have assumed that the particle is small enough that it does
not break the rotational symmetry about the axis of the cavity sufficiently to resolve the several degenerate
Hermite-Gaussian modes that superpose to yield the trapping Laguerre-Gaussian mode.

The interaction energy is minimized at $r=R, z=0$; the dielectric particle is
therefore trapped on this ring, which corresponds to a field maximum. The presence of the particle at the antinode
of the optical trapping field will shift the cavity resonance. However, this shift can be easily compensated by changing
the trap laser frequency by a small amount. A large $P_{in}'$ is chosen such that the trapping along the radial $(r)$ and
axial $(z)$ directions is stiff. The
corresponding particle displacements are then small and effectively decouple from the particle's azimuthal motion \cite{Isart2010}.
Furthermore, such small displacements may only pull the trap mode frequency by an amount which is much smaller
than the cavity linewidth, and is thus negligible \cite{Aspelmeyer2013}.
In the remainder of the article, we therefore ignore the axial and radial motion of the dielectric. Further, the intense
trapping optical field will be considered to be a parametric quantity and not a dynamical variable.

The azimuthal motion of the particle is free, and stable if the rotational energy
of the particle does not exceed the depth of the confining radial well. We will assume that the dielectric particle
experiences a constant torque $\tau$ causing it to rotate around the cavity axis. This torque can be enforced by
magnetic actuators if the particle is chosen to be a paramagnetic bead \cite{Sacconi2001}, or if its surface is partially
coated with metal, making it a ``dot Janus" particle \cite{Clark2010}. Such rotation techniques
have been demonstrated experimentally, albeit in liquid media; however, several experiments have also recently trapped and
rotated microparticles in free space \cite{Yoshi2013,Kane2010}. Rather than consider in further detail the mechanism
of rotation, which may be implemented in a number of ways, we proceed to focus instead on our detection scheme.

\subsection{Probe mode}
The rotation of the particle is weakly probed by a second optical field of wavelength $\lambda_{p} \neq \lambda_{t}$,
near-resonant with the cavity, and consisting of a superposition of two degenerate counter-rotating Laguerre-Gaussian
beams $LG_{\pm l}$ with orbital angular momenta $+l$ and $-l$ respectively. Near the trapping ring, the corresponding
mode function $\psi_{l}(\bold{r})$ obeys \cite{Isart2010}
\begin{equation}
\label{eq:LGLattice}
\left|\psi_{l}(\bold{r})\right|^{2}=\frac{1}{\left|l\right|!}
\left(\frac{R\sqrt{2}}{\omega_{0}}\right)^{\left|l\right|}e^{-\frac{2R^{2}}{\omega_{0}^{2}}}
\cos^{2}\left(k_{p}z\right)\cos^{2}\left(l\phi\right),
\end{equation}
where
$\omega_{0} \gg \lambda_{p}=2\pi/k_{p}$. In practice, the polarizations of the trapping $(\lambda_{t})$ and probe
$(\lambda_{p})$ fields may be chosen orthogonal, in order to eliminate interference effects. As can be seen from
Eq.~(\ref{eq:LGLattice}), the effect of the $LG_{\pm l}$ beams is to create an angular lattice with
$2l$ sites on the trapping ring (Fig.~\ref{fig:P1} displays the case $l=2$). This lattice breaks the continuous
azimuthal symmetry about the cavity axis and can thus sense the rotation of the particle \cite{Lavery2013,Phillips2014}, as
we show below.
\subsection{Optorotational coupling}
We now characterize the light-matter coupling for our system, using a classical model. A fully quantum mechanical
treatment will be the subject of later work. We denote the cavity
resonance frequency of the probe $LG_{\pm l}$ modes in the absence of the dielectric as $\omega_{c}$. In the presence of the
dielectric, this frequency shifts in a way that depends on the angular position $\phi$ of the dielectric \cite{Isart2010}.
Since $r_{d}<(\lambda_{p}, L)$, where $L$ is the length of the cavity, the shifted frequency $\omega(\phi)$ can be
evaluated using the perturbation-theoretic Bethe-Schwinger formula \cite{NovotnyBook}
\begin{equation}
\label{eq:BS}
\frac{\omega(\phi)}{\omega_{c}}=1-\frac{\int_V (\epsilon_{r}-1)\left|\psi_{l}(\bold{r})\right|^{2}d\bold{r}}
{\int_{V_{c}} \left|\psi_{l}(\bold{r})\right|^{2}d\bold{r}},
\end{equation}
where $V_{c}$ is the cavity mode volume. Using Eq.~(\ref{eq:LGLattice}) in the formula ~(\ref{eq:BS}) gives
\begin{equation}
\omega(\phi)=\omega_{c}-g(l)\cos^{2}l\phi,
\end{equation}
with
\begin{equation}
g(l) = \left(\epsilon_{r}-1\right)\frac{2^{\frac{l+3}{2}}}
{\Gamma \left(\frac{l+1}{2}\right)}
\left(\frac{R}{\omega_{0}}\right)^{l}\left(\frac{V}{\pi \omega_{0}^{2}L}\right)e^{-2\left(\frac{R}{\omega_{0}}\right)^{2}}\omega_{c},
\end{equation}
where $\Gamma [(l+1)/2]$ is a Gamma function. We note that in contrast to previous work \cite{Deych2011}, the dielectric
particle is far away from the cavity mirrors, and thus no appreciable rearrangement of charges occurs on the cavity walls,
and the standard form of the optomechanical coupling applies.
\section{Equations of motion}
To describe the dynamics of the optical field, we use the classical variable $a(t)$, such that
$\left|a(t)\right|^{2}$ is the number of photons in the cavity at time $t$. In order to characterize the rotor, we use the
classical angular momentum $L_{z}$ and the explicitly periodic angular displacement variable
\begin{equation}
U_{l}=e^{i2l\phi},
\end{equation}
where $\phi$ takes on continuous values. Using the variables just described, the coupled-mode equations of motion in
the frame of the laser driving the cavity with the $LG_{\pm}$ fields are
\begin{eqnarray}
\label{eq:OpticalFieldEvolution}
\dot{a}&=&\left\{i\left[\Delta'-\frac{g(l)}{2}\left(U_{l}+U_{l}^{*}\right)\right]-\frac{\gamma}{2}\right\}a+\sqrt{\gamma}a_{in},\\
\label{eq:AngleEvolution}
\dot{U_{l}}&=&\frac{i 2 l U_{l} L_{z}}{I},\\
\label{eq:AngularMomentumEvolution}
\dot{L_{z}}&=&-\gamma_{m}L_{z}-2il\hbar g(l)\left(U_{l}-U_{l}^{*}\right)\left|a(t)\right|^{2}+\tau+\tau_{in}.\\
\nonumber
\end{eqnarray}
We now describe each of the Eqs.(\ref{eq:OpticalFieldEvolution})-(\ref{eq:AngularMomentumEvolution}).
In Eq.~(\ref{eq:OpticalFieldEvolution}), which has been derived using the formalism of Haus \cite{HausBook},
\begin{equation}
\Delta'=\Delta-\frac{g(l)}{2},
\end{equation}
where $\Delta=\omega_{d}-\omega_{c}$ is the detuning
between the frequency $\omega_{d}$ of the $LG_{\pm l}$ driving laser and $\omega_{c}$,
$\gamma$ is the cavity loss rate, and $a_{in}=\sqrt{P_{in}/\hbar\omega_{c}}$, where $P_{in}$ is the input
power. Classical laser noise has been neglected, as well as the thermal
noise contribution to the radiation mode, which is very small at optical frequencies. Vacuum fluctuations in
the optical mode have also been ignored (but see below). Comparing Eq.~(\ref{eq:OpticalFieldEvolution}) to the
analogous equation for vibrational optomechanics immediately clarifies that the phase of the optical field is
sensitive to the $(U_{l}+U_{l}^{*})/2=\cos 2 l\phi$ quadrature of the mechanical rotation, and that this
sensitivity improves linearly with the cavity finesse \cite{Aspelmeyer2013}.

In Eq.~(\ref{eq:AngleEvolution}),
$I=m R^{2}$ is the dielectric particle's moment of inertia about the axis of rotation, and we have
implicitly used the relation $\dot{\phi}=L_{z}/I$ when taking the time derivative of $U_{l}$.

Equation ~(\ref{eq:AngularMomentumEvolution}) describes a rotor damped at a rate
$\gamma_{m}$, and $\tau_{in}$ is a Langevin torque with zero mean and the two-time fluctuation correlation
\begin{equation}
\langle \delta \tau_{in}(t)\delta \tau_{in}(t')\rangle=2I \gamma_{m}k_{B}T\delta(t-t'),
\end{equation}
signifying white noise. Unless
mentioned otherwise, we will assume that the externally applied torque is larger than the torque due to the optical lattice,
i.e. $\tau > 2il \hbar g(l)\left(U_{l}-U_{l}^{*}\right)\left|a(t)\right|^{2}.$
\subsection{Steady state}
The steady-state of the system can be found readily by equating the time derivatives in Eqs.~(\ref{eq:OpticalFieldEvolution})-(\ref{eq:AngularMomentumEvolution}) to zero,
\begin{equation}
\label{eq:SteadyState}
\omega_{m,s}=\frac{L_{z,s}}{I}=\frac{\tau}{I\gamma_{m}},
\end{equation}
\begin{equation}
\label{eq:SteadyState2}
U_{l,s}=0,
\end{equation}
\begin{equation}
\label{eq:SteadyState3}
a_{s}=\frac{\sqrt{\gamma}a_{in}}{\left[\Delta'^{2}+\left(\frac{\gamma}{2}\right)^{2}\right]^{1/2}}.
\end{equation}
Since the particle is in a non-equilibrium steady state, Equations ~(\ref{eq:SteadyState})-(\ref{eq:SteadyState3})
are to be interpreted as statements about time averages. For example Eq.~(\ref{eq:SteadyState2}) can be re-written as
\begin{eqnarray}
U_{l,s}&=&\langle U_{l}\rangle,\nonumber\\
&=&\langle e^{i2l\phi}\rangle,\nonumber\\
&=&\langle \cos 2l\phi + i\sin 2l\phi \rangle \nonumber\\
&=&\langle \cos 2l\phi \rangle + i \langle \sin 2l\phi \rangle,\\
\nonumber
\end{eqnarray}
where the brackets denote an average over one period of mechanical rotation. Thus, while $e^{il\phi}$ can never
equal zero as a function, it \textit{can} vanish \textit{on average}, since the cosine and sine average to zero
during one rotation.

Importantly, from Equation ~(\ref{eq:SteadyState}) it can be seen that the torque due to the probe lattice makes no
contribution to the average rotation rate $\omega_{m,s}$. This is because the lattice accelerates the particle over
half of its motion while decelerating it (almost) equally over the other half. This suggests that a rotation sensing
(velocimetry) scheme based on our proposal can circumvent, at least to lowest order, the effect of the optical probe
on the mechanical motion, and thus provide an accurate measurement of the undisturbed rotation frequency. This
statement will be supported both analytically as well as numerically below.

It is also worth noting that working directly
with $\dot{\phi}=L_{z}/I$ instead of Eq.~(\ref{eq:AngleEvolution}) would have implied no rotation in the steady
state, which is unphysical, since we have assumed that the external torque $\tau$ always overcomes the weak angular
lattice.
\subsection{Linear response}
We now examine the linear response of the system, by writing each dynamical variable as a sum of its steady state
value and a fluctuation \cite{Aspelmeyer2013}, e.g.
\begin{equation}
a=a_{s}+\delta a,
\end{equation}
Using   Eqs.~(\ref{eq:OpticalFieldEvolution})-(\ref{eq:SteadyState3})
and retaining only terms linear in the fluctuations, we find
\begin{eqnarray}
\label{eq:linearresponse1}
\dot{\delta a}&=&-i \frac{g(l)}{2}a_{s}\left(\delta U_{l}+\delta U_{l}^{*}\right)+\left(i\Delta'-\frac{\gamma}{2}\right)\delta a,\nonumber \\
\label{eq:linearresponse2}
\dot{\delta U_{l}}&=&\frac{i 2 l}{I}L_{z,s}\delta U_{l},\\
\label{eq:linearresponse3}
\dot{\delta L_{z}}&=&-\gamma_{m}\delta L_{z}+2il \hbar g(l)\left|a_{s}\right|^{2}\left(\delta U_{l}-\delta U_{l}^{*}\right)+\delta\tau_{in}.\nonumber\\
\nonumber
\end{eqnarray}
This set of linear first-order differential equations can be readily solved analytically for any initial
conditions $\delta a(0),\delta U_{l}(0)$ and $\delta L_{z}(0)$. The full solution is rather involved; here we
consider only a revealing limit: on resonance $(\Delta'=0)$, and ignoring cavity losses $(\gamma=0)$, we
find the Fourier transform of $\delta a(t)$ is
\begin{equation}
\label{eq:SidebandSpectrum1}
\delta a\left(\omega\right)= A \delta\left(\omega\right)
+B^{*}\delta\left(\omega-\omega_{s}\right)-B\delta\left(\omega+\omega_{s}\right),
\end{equation}
where
\begin{equation}
A=\sqrt{2\pi}\delta a(0)+B-B^{*},
\end{equation}
\begin{equation}
B=\frac{\sqrt{2\pi}a_{s}g(l)\delta U_{l}(0)}{\omega_{m,s}},
\end{equation}
and
\begin{equation}
\label{eq:SidebandFrequency}
\omega_{s}=2l\omega_{m,s}.
\end{equation}
The Dirac delta at $\omega=0$ on the right hand side of Eq.~(\ref{eq:SidebandSpectrum1}) corresponds to the optical
frequency, since we are in the frame rotating at the cavity resonance. Importantly, the Dirac deltas at $\omega=\pm \omega_{s}$
correspond to sidebands which are created as the particle rotating at the frequency $\omega_{m,s}$ encounters $2l$
optical lattice sites.
These sidebands fundamentally arise from the rotational Doppler shift
\cite{Courtial1998,Lavery2013} imprinted on the cavity photons by the mechanical motion, and are analogous to the sidebands
at the harmonic oscillator frequency seen in standard vibrational optomechanics experiments \cite{Aspelmeyer2013}.
The rotation frequency peak at $\omega_{s}$ can thus be recovered by homodyning the probe beam $LG_{\pm l}$ once it has exited the
cavity \cite{Aspelmeyer2013}. Interestingly, $l$ maybe treated as an adjustable parameter by the experimentalist to
allow convenient detection for a given rate of rotation.

Equation ~(\ref{eq:SidebandSpectrum1}) shows that the linear response is sensitive to the average mechanical
rotation rate $\omega_{m,s}$, and is insensitive to the action of the probe field on the mechanical motion.
It can readily be verified that this conclusion holds true even in the presence of detuning $(\Delta' \neq 0)$ and
optical damping $(\gamma \neq 0)$.
\subsection{Dynamics}
To analyze more comprehensively the dynamics of the system, we now numerically simulate
the full nonlinear model [Eqs.~(\ref{eq:OpticalFieldEvolution})-(\ref{eq:AngularMomentumEvolution})].
\subsubsection{Simulation parameters}
\label{subsubsec:Parameters}
To make
the simulation realistic, we include vacuum optical noise by using
\begin{equation}
\sqrt{\gamma}a_{in} \rightarrow \sqrt{\gamma}a_{in}+\sqrt{\gamma}\delta a_{in},
\end{equation}
in Eq.~~(\ref{eq:OpticalFieldEvolution}), with \cite{Aspelmeyer2013}
\begin{equation}
\langle \delta a_{in}(t)\rangle = 0,
\end{equation}
and
\begin{equation}
\langle\delta a_{in}(t)\delta a_{in}^{\dagger}(t')\rangle=\delta(t-t'),
\end{equation}
All stochastic processes are modeled as Gaussian white noises, and each simulation run includes a single
realization of the noise.

For the cavity, we assume $L=25$ cm, and $\mathcal{T}=8\times 10^{-4},$ which lead to $\gamma \simeq 150$ KHz.
The radius of curvature of both mirrors is assumed to be $40$m.
For the trapping field, $\lambda_{t}=1064$ nm, leading to $\omega_{0}=2$ mm, and $l'=2$, implying $R \simeq 2$ mm;
also, we assume $P_{in}'=180$ mW.
%Since the dielectric is trapped far $(\sim 0.5)$ cm from the mirrors, it does not experience any
%nonconservative couplings unlike in previous proposals, where the nanoparticle moves in the evanescent field of the optical
%resonator \cite{Deych2011}. We have also verified that the terms coupling the axial and radial motions to the rotation are
%negligible for the assumed trap power.
For the probe field, we assume $\lambda=980$ nm, $\Delta'=0, l=2$, yielding
$\omega_{c}\simeq 10^{15}$ Hz. The dielectric particle is chosen to be a polystyrene $(\epsilon_{r}=2.5)$ sphere of radius $
r_{d}=150$ nm. This gives $\alpha=1.2\times 10^{-31} $m$^{3}$ and $g(l=2)\simeq 25$ mHz. Using the polystyrene density $\rho=1050$
kg m$^{-3}$, we find $m=1.5\times 10^{-19}$ kg, and $I=62.5\times 10^{-24}$ kg m$^{2}$. From earlier work,
$\gamma_{m}\simeq 60$ Hz at a background pressure of $\sim 10$ mBar \cite{Novotny2012}; also, we assume
$\tau = 2.5$ fN$\mu$m, which is within demonstrated capabilities \cite{Clark2010}. These assumptions lead to
$L_{z,s}=8.3\times 10^{-23} $kg m$^{2}$s$^{-1}\sim 10^{10}\hbar$, justifying our classical treatment of the rotor, and
implying a rotational kinetic energy of $L_{z,s}^{2}/2I \simeq 107$ K. This is safely lower than the radial trapping well
depth $\simeq 360$ K; thus the particle remains bound as it rotates. The optical intensity experienced by
the particle is about $120$ kW cm$^{-2},$ several orders of magnitude below the threshold for heating due to absorption
\cite{Chang2010}. Since $r_{d} \ll \lambda_{p}$, the effect of recoil heating on
the particle motion is negligible \cite{Chang2010,Kiesel2013}; also we find that the scattering-induced photon loss
rate from the cavity is small ($\sim 30$Hz) \cite{Chang2010} compared to $\gamma$ calculated above and does not greatly
affect the cavity finesse.
\begin{figure}
\includegraphics[width=1.0\columnwidth]{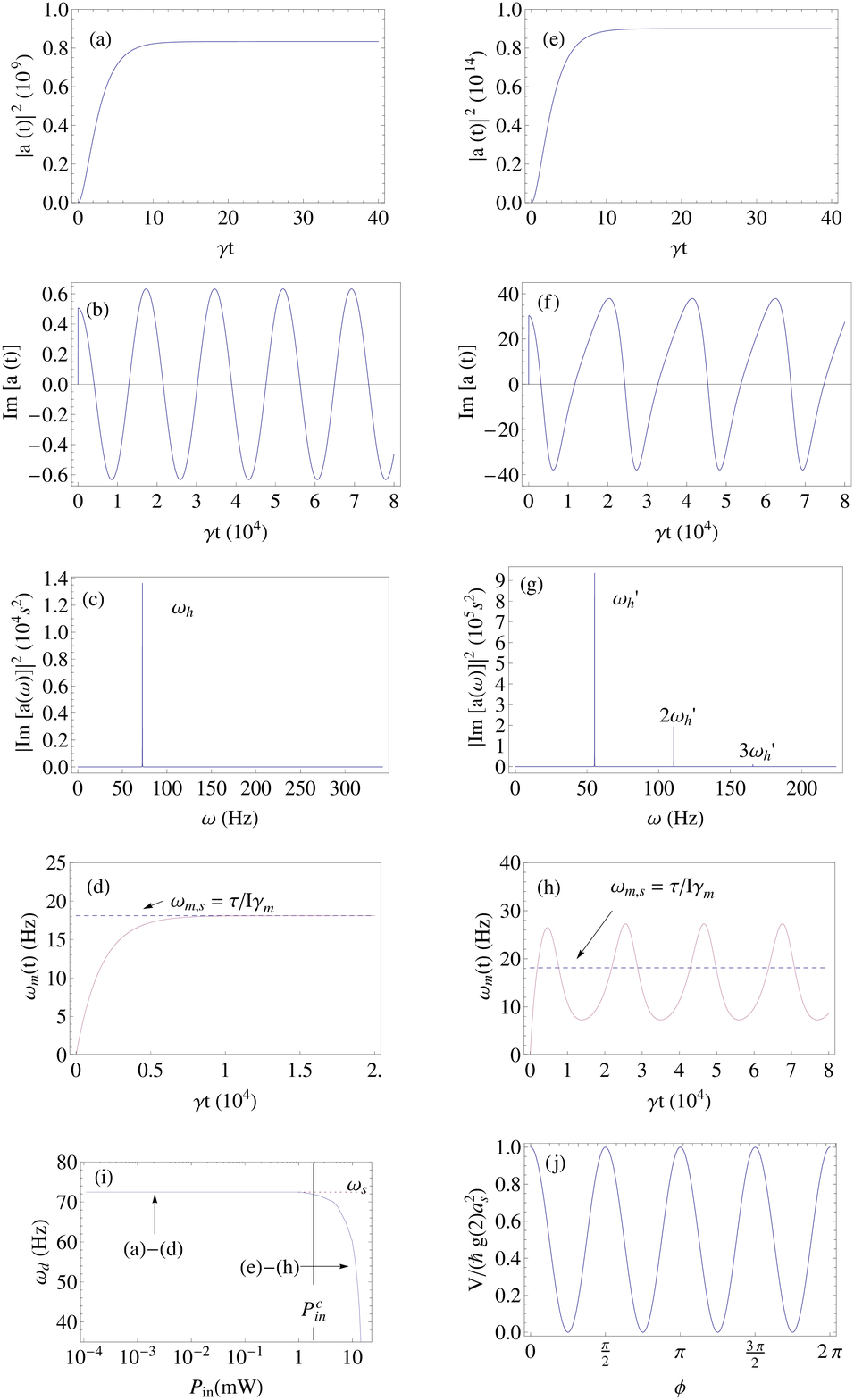}
\caption{Dynamics of the proposed optomechanical system for $T=25$mK (other parameters are stated in the text)
and various regimes of the probe power: (a)-(d) $P_{in}=2\, \mu$W, (e)-(h) $P_{in}=10 $ mW. (i) Homodyne peak
location versus probe power. (j) Potential energy of the optical lattice. The plots are described in detail in
the text.}
\label{fig:P2}
\end{figure}
\subsubsection{Dynamics: linear regime}
Using the standard methods of dynamical systems theory, we have ensured that all the configurations discussed
in this article are mechanically stable \cite{Genes2008}.
Starting from a cavity empty of probe radiation, and a particle at rest at $\phi = 37\,^{\circ}$,
we now consider the behavior of the system for various probe powers $P_{in}$.
In Fig.~\ref{fig:P2} (a)-(d), $P_{in} = 2 \mu$W, which lies in the regime of linear response, but is adequate for
carrying out homodyne detection. Figure (a) shows
the number of probe photons in the cavity reaching an equilibrium value; (b) shows that the imaginary part of the
cavity field, which can be measured using homodyne detection, displays sinusoidal oscillations due to mechanical
modulation; (c) shows the Fourier transform of (b) with a frequency peak at $\omega_{h}=72.463761 \pm 10^{-6}$Hz, which
accurately reflects the nominal value $\omega_{s}$ from Eq.(\ref{eq:SidebandFrequency})
(d) shows the time dependent mechanical rotation rate $\omega_{m}(t)$ rising to the steady state value
$\omega_{m,s}=\omega_{s}/4$. Further numerical exploration shows that the limits on the frequency resolution
arise mostly from thermal and to a lesser extent from photonic noise.
\subsubsection{Dynamics: nonlinear regime}
In Figs.~\ref{fig:P2} (e)-(h), $P_{in}=10$ mW, which lies in the regime of the system's nonlinear response, i.e.
when the variables deviate appreciably from their steady state values.
Figure (e) is similar to (a); (f) however shows a more complicated time evolution than (b) and suggests the
presence of multiple underlying Fourier frequencies; indeed, (g) shows a peak at $\omega_{h}'=55.26934 \pm 0.00002$Hz,
and higher sidebands at integer multiples of $\omega_{h}'$; (h) shows large optical modulation of the mechanical
rotation rate and may be compared to (d).

\subsubsection{Velocimetry}
The two system response regimes, linear as well as nonlinear, are summarized together in Fig.~\ref{fig:P2}(i), which
shows the homodyne-detected frequency $\omega_{d}$ versus the probe mode power $P_{in}$. In the linear response regime of low probe
power, as $P_{in}$ is varied, the homodyne signal frequency $\omega_{d}$ stays constant and close to the
nominal value $\omega_{s}$. The identification of this regime of accurate response is the central result of this article.
In this regime, the potential energy of the lattice [Fig.~\ref{fig:P2}(j)] is much smaller
than the kinetic energy of the particle, and modifies the rotation frequency very little from the nominal value
$\omega_{m,s}$. While Figs.~(\ref{fig:P2})(a)-(d) display the linear regime results for $T=25$ mK, runs for $T=300$ K
[not shown in Fig.~(\ref{fig:P2})] yield $\omega_{h}=72.4738\pm 10^{-4}$ Hz. These measurements are of at least an
order of magnitude higher resolution than state-of-the-art velocimetry experiments
\cite{Chavez2002,Kane2010,Yoshi2013,Lavery2013,Phillips2014},
implying that cavity-based rotation sensing may be a direction worth exploring for rotational Doppler velocimetry.
While we have investigated the detection of low angular velocities (few tens of Hz) for simplicity, in practice it might be
more convenient to work with faster rotation rates, due to the presence of low frequency noise as well as fluctuations in background
pressure and the applied torque $\tau$.

In the nonlinear regime of high probe power [beyond the critical power $P_{in}^{c}$, supplied as a guide to the eye,
in Fig.~(\ref{fig:P2})(i)], the particle kinetic and lattice potential energy are comparable, and the negotiation of the
lattice adds a large delay to the orbital period of the particle. This lowers the average particle rotation rate significantly from its
unperturbed value $\omega_{m,s}$, which can no longer be measured accurately. Eventually, for very large $P_{in}$, the
particle is trapped in one of the lattice minima.

\section{Conclusion}
We have theoretically explored the optomechanical coupling of nanorotation to a cavity mode carrying orbital
angular momentum. Our results suggest that optical cavities deserve to be investigated further as a means of advancing
the technique of rotational Doppler velocimetry.

Our analysis can be extended to the case $\tau = 0,$ in which case the
only external torque present is thermal, and the dielectric particle travels on the angular periodic lattice created by
the probe mode. This would correspond to Brownian motion in a periodic angular lattice, which is of interest, for example,
to studies of dielectric relaxation \cite{CoffeyBook}. This would likely require very low background pressures. It would
also be interesting to generalize our classical treatment to the quantum regime, to determine, for example, what the
minimum detectable rotation rate - i.e. mechanical angular momentum - is. Lastly, our analysis may be extended
to describe the motion of an atom in place of the dielectric, as the optomechanical coupling will be of the same
functional form in the angular coordinate, if the probe laser is detuned from an atomic resonance, and the
interaction is consequently due to a dipole potential.

We are grateful to S. Agarwal, B. Zwickl, M. Vengalattore, H. Shi, B. Ek and B. Rodenburg for useful discussions,
and to the Research Corporation for Science Advancement for support.

\end{document}